# AGGREGATION RATES IN ONE-DIMENSIONAL STOCHASTIC SYSTEMS WITH ADHESION AND GRAVITATION

By Mikhail Lifshits[1] and Zhan Shi

*St. Petersburg State University, Université Lille I and Université Paris VI*

We consider one-dimensional systems of self-gravitating sticky particles with random initial data and describe the process of aggregation in terms of the largest cluster size $L_n$ at any fixed time prior to the critical time. The asymptotic behavior of $L_n$ is also analyzed for sequences of times tending to the critical time. A phenomenon of phase transition shows up, namely, for small initial particle speeds ("cold" gas) $L_n$ has logarithmic order of growth while higher speeds ("warm" gas) yield polynomial rates for $L_n$.

**1. Introduction.** We consider a system of $n$ particles living in one-dimensional space. At initial moment, every particle is characterized by its mass, initial position and initial speed. There exists a pairwise gravitation between the particles. Between the moments of shocks, particles move in this gravitation field according to usual rules of Newtonian mechanics. The shock between two (or more) particles results in the birth of a new particle ("cluster") whose characteristics are defined by the laws of conservation of mass and momentum (while the energy is dissipated at these nonelastic collisions).

The clusters gradually become larger and larger while the number of clusters diminishes—until the unique cluster, containing the totality of mass, remains on the line. This collapse vaguely models emergence of a "star" from dispersed "dust." Indeed, the roots of the model are in astrophysics (see [14, 18]), but the reason of actual interest in similar particle systems is due to their relation with solutions of nonlinear PDEs such as the Burgers equation (see [1, 3, 4, 7] and the references therein).

The aim of the present work is to describe essential features of the aggregation process provided that initial data are *random* and that the number

Received June 2003; revised March 2004.
[1]Supported in part by Grants RFBR 02-01-00265 and NSh-2258.2003.1.

of initial particles tends to infinity. Assuming that at the beginning there exist $n$ particles, we try to understand the behavior of $L_n(t)$—the largest number of particles that form a common cluster at time $t$. Therefore, $L_n(t)$ increases, as time goes by, from 1 at the beginning to $n$ at the collapse time.

We will extend the quantitative results obtained in recent works [[2, 7, 10], [16, 17]]. It is known from these works that, under reasonable assumptions, there exists a *critical time* $T^*$ such that $\lim_{n\to\infty} \frac{L_n(t)}{n} = 0$ in probability for every $t < T^*$ and $\lim_{n\to\infty} \frac{L_n(t)}{n} = 1$ in probability for every $t > T^*$. In other words, the time when essential collapse occurs, is, in fact, deterministic. "Essential" means that after critical time one can still observe a number of small peripheral clusters, but the main core has already been formed and it contains the overwhelming part of the total mass.

Our results show that the aggregation process behaves rather differently in the cases of "large" and "small" initial speeds. In the literature the most frequently used model is that of i.i.d. initial speeds. For this case, which is naturally interpreted as "warm" system (see, e.g., [10, 17]), we essentially show that for any $t < T^*$,

$$L_n(t) \approx c(t) n^{2/3} (\log n)^{1/3}.$$

On the other hand, whenever initial speed is small or vanishes, which is natural to interpret as "cold" system (see [7]), one has

$$L_n(t) \approx c(t) \log n.$$

Our results, in fact, cover the whole range of possible behaviors containing the two aforementioned special cases.

In any case, one can observe from these formulae that $L_n(t) \ll n$, that is, the aggregation process is rather slow and that genuinely macroscopic clusters appear only shortly before the critical time. Therefore, it is interesting to describe the behavior of $L_n(t_n)$ for $t_n \to T^*$. This is done for cold systems in Theorem 3.4.

We give a rigorous description of the model in Section 2 and state our results in Section 3. Section 4 provides the necessary information on conservation laws which control the behavior of the systems. Finally, Section 5 contains the collection of proofs. There is an amazing contrast between the elementary nature of the results and the rather advanced techniques one needs to obtain them.

## 2. Systems of sticky particles.

2.1. *Dynamics of deterministic systems.* We consider the following one-dimensional system (gas) of particles with gravitation. At the starting moment $t = 0$, our gas consists of $n$ particles, positioned on the real line at points with coordinates $x_1(0), \ldots, x_n(0)$. Particles are always enumerated so that $x_1(0) \leq \cdots \leq x_n(0)$. Each particle is characterized by its mass $m_i$ and its initial speed $v_i(0)$. The particles move under the action of forces of pairwise mutual gravitation. The gravitation force acting on a particle of mass $m$ positioned at $x$, from a particle of mass $\mu$ positioned at $y$, is

$$(2.1) \qquad F = \gamma m \mu \operatorname{sign}(y - x).$$

Here, $\gamma > 0$ is a positive *gravitation constant*. Note that there is no dependence of the gravitation force on the distance between the particles. This feature is typical for one-dimensional gravitation models.

Between collisions particles obey the Newton's second law ($F = m\frac{d^2x}{dt^2}$). The total gravitation force acting on a particle is obviously proportional to the difference of the masses on its right and on its left.

On collision particles stick together, following the conservation laws of mass and momentum. In other words, two particles with characteristics $(m, v)$ and $(\mu, w)$ produce one particle with mass $M = m + \mu$ and speed $V = (mv + \mu w)/M$. This is a completely nonelastic collision, kinetic energy is dissipated. The particles which are born upon collisions are called *clusters*.

We assume that the initial particles do not die at collisions but continue their movement as parts of the created clusters. Therefore, the position, the speed and the acceleration of a particle are understood as those of the cluster containing this particle. We denote them by $x_i(t)$, $v_i(t) := x_i'(t)$ and $x_i''(t)$, respectively, for each $t \geq 0$. The destiny of each particle is therefore defined during the life time of the system. The speed and the acceleration have jumps at the times of collisions of the particle or of the cluster which contains it (actually, the acceleration also has $\delta$-s at collision times).

Throughout the paper, we call *particles* only the initial particles, and *clusters* the products of collisions as well as the initial particles. Thus, at any time, the system consists of a number of clusters, each cluster being a set of one or more particles.

When $v_i(0) = 0$ (zero initial speed), the system is referred to as a *cold gas*. If a gas is not cold, we call it a *warm gas*.

2.2. *Similarity of systems.* Let positive numbers (similarity coefficients) $c_x$, $c_v$, $c_m$, $c_t$ and $c_\gamma$ satisfy the conditions

$$1 = \frac{c_x}{c_v c_t} = \frac{c_v}{c_m c_\gamma c_t}.$$

Consider two systems of particles. The first system has initial data $(x_i(0), v_i(0), m_i)$ and gravitation coefficient $\gamma$, while the second has initial data $(c_x x_i(0), c_v v_i(0), c_m m_i)$ and gravitation coefficient $c_\gamma \gamma$. These two systems are *similar* at any time $t \geq 0$, that is, the first system at time $t$ has at position $x$ a cluster of mass $m$ moving with the speed $v$ if and only if the second system at time $c_t t$ has at position $c_x x$ a cluster of mass $c_m m$ moving with the speed $c_v v$.

2.3. *Stochastic systems of particles.* Let us introduce a stochastic version of the problem by considering initial parameters of the particles as random variables, while the interaction rules and the dynamics remain deterministic and follow the rules described above. Moreover, we consider the asymptotic situation, in which the number $n$ of particles tends to infinity, while other parameters vary in a reasonably consistent way.

2.3.1. *Masses.* We assume that the masses of the particles are deterministic and equal to $\rho n^{-1}$ each, where $\rho > 0$ is a fixed constant. This leads, at the limit, to a model with a given density of the matter $\rho$. We call the *size* of a cluster the number of particles in it.

2.3.2. *Initial speeds.* We assume that the initial speeds of the particles have the form $v_i(0) = \sigma_n u_i, 1 \leq i \leq n$, where $(u_i)$ is a collection of i.i.d. random variables with zero mean and unit variance. Zero mean is assumed only for convenience of notation. It is easy to see that adding the same constant to all initial speeds affects neither the times of collisions nor the sizes of the created clusters. The only consequence is an additional uniform drift of the whole system.

In the literature, the scaling parameter $\sigma_n$ is usually assumed to be independent of $n$. Then the case $\sigma_n = 0$ corresponds to a cold gas, while $\sigma_n = \sigma > 0$ corresponds to a warm gas. We prefer to handle the setting with variable $\sigma_n$, since we are interested in identifying the border separating the models whose properties are close to those of cold and warm gas, respectively.

2.3.3. *Initial positions.* Three meaningful different models are distinguished here, though they often lead to the same asymptotic results.

The *lattice deterministic model* assumes that the particles are initially located on the lattice $x_i(0) = i/n$, $1 \leq i \leq n$.

The *Poisson model* assumes that the particles are initially located on the positive half-line at the first $n$ points of a Poisson point process with intensity $n$. In other words, they are located at the times of first jumps of a Poisson process of the just mentioned intensity (the space where the particles live is interpreted here as a time parameter of the process). By the well-known property of Poisson processes, the differences $x_i(0) - x_{i-1}(0)$ [notation: $x_0(0) := 0$], $1 \leq i \leq n$, are independent exponential random variables with mean $n^{-1}$.

The *i.i.d.-model* assumes that the particles are initially located at the points corresponding to a sample of $n$ independent random variables uniformly distributed on the interval $[0, 1]$. Since the particles are indexed by the order of initial positions, the initial location of the $i$th particle corresponds to the $i$th order statistics of the uniform sample.

2.3.4. *Relation between the Poisson and the i.i.d.-models.* The Poisson model is more convenient for investigation due to the (aforementioned) property of independence of distances between the particles at time zero. The following passage from the Poisson model to the i.i.d.-model is useful; it is well known and stated here without proof.

FACT 2.1. Fix $k \geq 1$, and let $X_1, \ldots, X_k, X_{k+1}$ be the times of first $k$ jumps of a Poisson process of constant intensity. Then the random variables $\widetilde{X}_i = \frac{X_i}{X_{k+1}}$, $1 \leq i \leq k$, have the same joint distribution as the order statistics of a sample of $k$ i.i.d. random variables uniformly distributed on the interval $[0, 1]$. Moreover, the random vector $(\widetilde{X}_1, \ldots, \widetilde{X}_k)$ and the random variable $X_{k+1}$ are independent.

Note that the intensity of the Poisson process does not play any role. In the sequel we use a standard Poisson process (of unit intensity), as well as a Poisson process of intensity $n$.

**3. Main results on aggregation rates.** In this section we investigate the asymptotic behavior of $L_n(t)$, the largest cluster size at time $t$ in the system of $n$ initial particles. Since the cluster size coincides with the number of initial particles in it, we always have $1 \leq L_n(t) \leq n$.

The *collapse time* $T(n)$ is the first time when all the particles belong to a common cluster, that is, when the total collapse of the particle system occurs.

Let us introduce the *critical time of the system* defined by

$$(3.1) \qquad T^* := (\gamma \rho)^{-1/2}.$$

3.1. *Aggregations in a cold gas.* First we recall a known result (see, e.g., [7]) on the collapse time, which motivates the definition (3.1).

FACT 3.1. *In a cold gas, for any of the three models (lattice, Poisson or i.i.d.) of initial positions, the random variable $T(n)$ converges in probability (when $n \to \infty$) to the deterministic constant $T^*$.*

Therefore, the behavior of the largest cluster size after the critical time is trivial:

COROLLARY 3.2. *In a cold gas, for any of the three models (lattice, Poisson or i.i.d.), for any $t > T^*$, we have $\lim_{n \to \infty} \mathbb{P}\{L_n(t) = n\} = 1$.*

We consider now $L_n(t)$ for $t < T^*$, that is, we are interested in the largest cluster size at a fixed time prior to the critical time.

The case of lattice initial positions is particularly instructive. For a cold gas we deal here with a purely deterministic model. All the particles move to the point of general meeting without any collision. More precisely, the particle $i$ moves along the trajectory

$$x_i(s) = \frac{i}{n} + \frac{\gamma \rho s^2}{2n}(n - 2i + 1),$$

with constant acceleration. We see that at time $T^*$ all the particles simultaneously meet at the barycenter point $x = \frac{n+1}{2n}$. The absence of collisions prior to the global meeting follows, for example, from the formula

$$x_i(s) - x_{i-1}(s) = \frac{1 - \gamma \rho s^2}{n}.$$

However, in other models of initial positions, the situation is not so trivial.

THEOREM 3.3. *In a cold gas for both Poisson and i.i.d.-models, for any $t < T^*$, we have*

(3.2) $$\lim_{n \to \infty} \frac{L_n(t)}{\log n} = I((t/T^*)^2)^{-1} \qquad \textit{in probability,}$$

*where*

(3.3) $$I(r) := r - 1 - \log r, \qquad r > 0.$$

The logarithmic order, already established in [7], indicates a very slow growth of clusters in a cold gas. Our contribution is to prove the existence of the limit and to determine its exact value.

The function $I(\cdot)$ in (3.3) is the rate function in the large deviations for the exponential law.

We set aside the delicate question of studying the behavior of $L_n(T^*)$, and consider the maximal cluster size in a cold gas at times shortly prior to the critical instant $T^*$.

THEOREM 3.4. *In a cold gas for both Poisson and i.i.d.-models, for any sequence $(t_n) \subset [0, T^*]$ which satisfies assumptions*

(3.4) $$\lim_{n \to \infty} t_n = T^* \quad \textit{and} \quad \lim_{n \to \infty} n(T^* - t_n)^2 = \infty,$$

*we have*

$$\lim_{n \to \infty} \frac{L_n(t_n)(T^* - t_n)^2}{\log(n(T^* - t_n)^2)} = \frac{(T^*)^2}{2} \qquad \textit{in probability.}$$

Assumption (3.4) is quite natural, since it is shown in [7] for total collapse time $T(n)$ that $T^* - T(n)$ has order $n^{-1/2}$. The assertion of our theorem obviously becomes false when $(t_n)$ approaches $T^*$ faster than what is admitted in (3.4).

3.2. *Aggregations in a warm gas.* Consider now a warm gas, that is, a system of particles with nonzero initial speeds $v_i(0) = \sigma_n u_i$, where the speed scale $(\sigma_n)$ is a sequence of nonnegative numbers and $(u_i)$ is a family of i.i.d. random variables with $\mathbb{E}(u_i) = 0$ and $\mathbb{E}(u_i^2) = 1$.

The aggregation in a warm gas has been extensively studied in [2, 10, 16, 17]. The critical time $T^*$ is still defined by (3.1), but its interpretation becomes slightly different. It is not interpreted any more as a limit value of collapse times, but as a time after which there exists a single cluster of huge mass and, possibly, a dust of small clusters whose total mass is negligible. In other words, analogously to Corollary 3.2, for any $t > T^*$, we have $\lim_{n\to\infty} \frac{L_n(t)}{n} = 1$ in probability.

The aggregation in a warm gas at critical time $T^*$ is remarkably studied in [17] (for lattice model with $\sigma_n = 1$). Therefore, it remains to evaluate $L_n(t)$ at times $t < T^*$ (cf. Theorem 3.2.1 in [17]).

Before stating the result, let us explain a natural but somewhat nonstandard notation of superior and inferior limits in probability which will be frequently used in the sequel. Let $\xi_n$ be a sequence of random variables and $c \in \mathbb{R}$. Then

$$\liminf_{n\to\infty} \xi_n \geq c \quad \text{in probability}$$

means that for all $\varepsilon > 0$,

$$\lim_{n\to\infty} \mathbb{P}(\xi_n \leq c - \varepsilon) = 0.$$

Similarly,

$$\limsup_{n\to\infty} \xi_n \leq c \quad \text{in probability}$$

means that for all $\varepsilon > 0$,

$$\lim_{n\to\infty} \mathbb{P}(\xi_n \geq c + \varepsilon) = 0.$$

THEOREM 3.5. *Assume that*

(3.5) $$\lim_{n\to\infty} \frac{\sigma_n}{n^{-1}\log n} = \infty \quad \text{and} \quad \lim_{n\to\infty} \frac{\sigma_n}{n^{1/2}} = 0,$$

*and that $u_i$ has a finite exponential moment $\mathbb{E}\exp\{a|u_i|\} < \infty$ for some $a > 0$. Then, for any of the three models (lattice, Poisson or i.i.d.), for any $t < T^*$, we have*

(3.6) $$\liminf_{n\to\infty} \frac{L_n(t)}{(n\sigma_n)^{2/3}(\log(n/\sigma_n^2))^{1/3}} \geq c_1(t) \quad \text{in probability,}$$

(3.7) $$\limsup_{n\to\infty} \frac{L_n(t)}{(n\sigma_n)^{2/3}(\log(n\sigma_n^{2/5}))^{1/3}} \leq c_2(t) \quad \text{in probability,}$$

*where*

(3.8) $$c_1(t) := 2\left(\frac{t}{1-\gamma\rho t^2}\right)^{2/3}, \qquad c_2(t) := (20/3)^{1/3} c_1(t).$$

In the basic case $\sigma_n = 1$, we get the estimate $L_n(t) \approx n^{2/3} (\log n)^{1/3}$.

For relatively large $\sigma_n$ (high temperature gas), a critical order turns out to be $\sigma_n \approx n^{1/2}$. When approaching this order, the main term in the asymptotic expression for $L_n(t)$ has order $n$, while the logarithmic terms of our upper and lower bounds are not of the same order anymore. This would indicate a kind of phase transition. The critical time for such a hot gas should not follow the formula $T^* = (\gamma\rho)^{-1/2}$, but tend to infinity.

It is interesting to understand what happens if we replace the assumption $\mathbb{E}\exp\{a|u_i|\} < \infty$ by the weaker condition

(3.9) $$\mathbb{E}[|u_i|^p] < \infty.$$

Then a small number of particles with high initial speeds can substantially perturb the behavior of $L_n(t)$. However, the following result says that if $p$ is large enough, then Theorem 3.5 still holds true, but in narrower zones of speed ranges.

PROPOSITION 3.6. *Assume that* (3.5) *and* (3.9) *hold. Let* $\varepsilon > 0$.

(a) *Whenever* $p > 2$ *and* $\sigma_n \geq \varepsilon n^{(3-p)/p}$, *we have* (3.6).
(b) *Whenever* $p > 6$ *and* $\sigma_n \geq \varepsilon n^{(7-p)/(p-4)}$, *we have* (3.7).

It is worthwhile to compare Proposition 3.6 with estimates in [17]. In our notation, Theorem 3.2.1 in [17] yields $L_n(t) \leq n^h$, $\forall h > \frac{2(p+2)}{3p}$, for $\sigma_n = 1$ and $p > 4$, while Proposition 3.6 provides a better bound $L_n(t) \leq \text{const} \times n^{2/3}(\log n)^{1/3}$. These bounds become closer when $p \to \infty$.

On the opposite side of the scale, for small $\sigma_n$ (we call it a low temperature gas), namely, for $\sigma_n \ll n^{-1} \log n$, it is easy to establish the same behavior as for the cold gas, that is, (3.2) is true.

Moreover, if we *formally* apply Theorem 3.5 to critically small $\sigma_n = \text{const} \times n^{-1} \log n$, we also get $L_n(t) \approx \log n$ just as in Theorem 3.3. More rigorously, we have the following analogue of Theorem 3.5.

THEOREM 3.7. *Assume that*

(3.10) $$\lim_{n\to\infty} \frac{\sigma_n}{n^{-1}\log n} = c \in (0,\infty),$$

*and that* $\mathbb{E}\exp\{a|u_i|\} < \infty$ *for some* $a > 0$. *Then, for any of the three models (lattice, Poisson or i.i.d.), for each* $t < T^*$,

$$\liminf_{n\to\infty} \frac{L_n(t)}{\log n} \geq \hat{c}_1(t) \qquad \text{in probability,}$$

$$\limsup_{n\to\infty} \frac{L_n(t)}{\log n} \leq \hat{c}_2(t) \qquad \text{in probability,}$$

where $0 < \hat{c}_1(t) \leq \hat{c}_2(t) < \infty$ are constants.

**4. Barycenter technique.** In this section we describe and extend a barycenter technique for the study of particle systems. This technique has been already used, for example, in [2, 6, 7, 10, 17].

We identify the particles with their numbers, and call a *block of particles* any set of particles which have consecutive numbers. We denote such a block by $J = (i, i+k]$; this block contains the particles numbered from $i+1$ to $i+k$.

The number $k$ is called the *size* of the block. It is worthwhile to mention that at any time $t > 0$, some particles in the block may belong to a cluster containing particles that are not in the block.

We say that a block is *free from the right* up to time $t$ if none of its particles has collided up to time $t$ with any of the particles initially located to the right of the block. A block *free from the left* is defined similarly. Finally, a block is *free* up to time $t$ if it is free both from the left and from the right. We note that collisions inside a free block are possible.

Let $M_J := \sum_{j \in J} m_j$. Define the *barycenter* of the block $J$ by $\bar{x}_J(t) := M_J^{-1} \sum_{j \in J} m_j x_j(t)$.

Define

$$(4.1) \qquad \bar{x}_J^*(s) := \bar{x}_J(0) + \bar{x}_J'(0)s + \tfrac{1}{2}\gamma(M_J^{(R)} - M_J^{(L)})s^2,$$

where $M_J^{(R)}$ and $M_J^{(L)}$ denote the total masses of the particles to the right and to the left of $J$, respectively. Note that $s \mapsto \bar{x}_J^*(s)$ represents the trajectory of the barycenter of the block $J$ without taking into account the collisions with external particles and is completely expressed in terms of initial data of the particles in the block $J$.

The following observation contains a basic idea in the study of such particle systems:

"The barycenter of a free block moves with constant acceleration, as if it were a single particle with mass equal to the total mass of the particles in the block."

More precisely, we have the following.

PROPOSITION 4.1. *Let a block $J$ be free from the right up to time $t$. Then*

$$(4.2) \qquad \bar{x}_J(s) \geq \bar{x}_J^*(s), \qquad s \in [0, t].$$

*Similarly, if a block $J$ is free from the left up to time $t$, then*

$$(4.3) \qquad \bar{x}_J(s) \leq \bar{x}_J^*(s), \qquad s \in [0, t].$$

The proof of Proposition 4.1 is straightforward, and is omitted. An immediate consequence is the following

COROLLARY 4.2. *Let a block $J$ be free up to time $t$. Then*

$$\bar{x}_J(s) = \bar{x}_J^*(s), \qquad 0 \le s \le t.$$

We call the *collapse time $t_J^{\text{cl}}$ of a block* the first time when all the particles of the block $J$ are in a common cluster. It is possible to express $t_J^{\text{cl}}$ via the random variables of type $\bar{x}_A^*; A \subset \{1,2,\ldots,n\}$. Toward this aim, let us introduce for every particle $j$, the first time $t_j^{\text{cl}}$ when it collides with the neighboring particle $j+1$. It is clear that for $J = (i, i+k]$, we have

$$(4.4) \qquad t_J^{\text{cl}} = \max_{0 < r < k} t_{i+r}^{\text{cl}}.$$

On the other hand, let us consider for each triplet $\alpha < j < \beta$, the quadratic function

$$Q_{\alpha,j,\beta}(s) := \bar{x}_{(\alpha,j]}^*(s) - \bar{x}_{(j,\beta]}^*(s).$$

We know that

$$Q_{\alpha,j,\beta}(0) \le x_j(0) - x_{j+1}(0) < 0.$$

Moreover, the main coefficient of the quadratic function is positive since by (4.1) it is equal to

$$\frac{\gamma}{2}[(M_{(\alpha,j]}^{(R)} - M_{(j,\beta]}^{(R)}) + (M_{(j,\beta]}^{(L)} - M_{(\alpha,j]}^{(L)})] = \gamma M_{(\alpha,\beta]} > 0.$$

Therefore, there exists a unique $\tau_{\alpha,j,\beta}^* > 0$ such that

$$Q_{\alpha,j,\beta}(\tau_{\alpha,j,\beta}^*) = 0.$$

From the mechanical point of view, $\tau_{\alpha,j,\beta}^*$ is the time when the trajectories of barycenters of blocks $(\alpha, j]$ and $(j, \beta]$ meet, as long as we do not take shocks into account. The value $\tau_{\alpha,j,\beta}^*$ has an explicit expression via the coefficients of the polynomial $Q_{i,j,\beta}$, in terms of initial positions, speeds and masses of the particles. It is worth noticing that for any $t > 0$,

$$(4.5) \qquad t \ge \tau_{\alpha,j,\beta}^* \iff Q_{\alpha,j,\beta}(t) \ge 0.$$

It turns out that the following is true.

PROPOSITION 4.3. *For any $j < n$, we have*

$$(4.6) \qquad t_j^{\text{cl}} = \min_{\alpha < j, \, \beta > j} \tau_{\alpha,j,\beta}^*.$$

PROOF. At any time $s < t_j^{\mathrm{cl}}$, the particles $j$ and $j+1$ belong to different clusters, so that any block of the form $(\alpha, j]$ is free from the right and any block of the form $(j, \beta]$ is free from the left. By Proposition 4.1,

$$\bar{x}^*_{(\alpha,j]}(s) \leq \bar{x}_{(\alpha,j]}(s) \leq x_j(s) < x_{j+1}(s) \leq \bar{x}_{(j,\beta]}(s) \leq \bar{x}^*_{(j,\beta]}(s).$$

Therefore, $Q_{\alpha,j,\beta}(s) < 0$. It follows that $t_j^{\mathrm{cl}} < \infty$ and by continuity of $Q$, we have $Q_{\alpha,j,\beta}(t_j^{\mathrm{cl}}) \leq 0$. Hence, $\tau^*_{\alpha,j,\beta} \geq t_j^{\mathrm{cl}}$. Since $\alpha < j$ and $\beta > j$ are arbitrary, we have $t_j^{\mathrm{cl}} \leq \min_{\alpha<j,\,\beta>j} \tau^*_{\alpha,j,\beta}$.

To prove the inequality in the other direction, we note that by definition, $t_j^{\mathrm{cl}}$ is the collision time between particles $j$ and $j+1$, so that there exist $\alpha < j$ and $\beta > j$ such that the blocks $(\alpha, j]$ and $(j, \beta]$ are free up to time $t_j^{\mathrm{cl}}$ (and collide at time $t_j^{\mathrm{cl}}$). By Corollary 4.2,

$$\bar{x}^*_{(\alpha,j]}(t_j^{\mathrm{cl}}) = \bar{x}_{(\alpha,j]}(t_j^{\mathrm{cl}}) = \bar{x}_{(j,\beta]}(t_j^{\mathrm{cl}}) = \bar{x}^*_{(j,\beta]}(t_j^{\mathrm{cl}}).$$

Hence, $t_j^{\mathrm{cl}} = \tau^*_{\alpha,j,\beta}$. It follows that $t_j^{\mathrm{cl}} \geq \min_{\alpha<j,\,\beta>j} \tau^*_{\alpha,j,\beta}$. The proposition is proved. $\square$

When estimating the collapse time of a block, it is sometimes more convenient to deal with simpler expressions. For example, it is a consequence of (4.4) and (4.6) that for $J := (i, i+k]$,

$$(4.7) \qquad t_J^{\mathrm{cl}} \leq \max_{0<r<k} \tau^*_{\alpha,i+r,\beta} \qquad \forall \alpha \leq i,\ \forall \beta \geq i+k.$$

In particular, taking $\alpha = i$ and $\beta = i+k$ gives

$$(4.8) \qquad t_J^{\mathrm{cl}} \leq \max_{0<r<k} \tau^*_{i,i+r,i+k}.$$

Note that the expression on the right-hand side depends only on the initial data of the particles in the block itself.

There is a situation when (4.8) is sharp.

PROPOSITION 4.4. *If a block $J = (i, i+k]$ is free up to its collapse time $t_J^{\mathrm{cl}}$, then*

$$(4.9) \qquad t_J^{\mathrm{cl}} = \max_{0<r<k} \tau^*_{i,i+r,i+k}.$$

PROOF. It is sufficient to prove the inequality opposite to (4.8). Take arbitrary integer $r \in [1, k)$ and consider the blocks $(i, i+r]$ and $(i+r, i+k]$. We know that, up to time $t_J^{\mathrm{cl}}$, the first block is free from the left, while the second is free from the right. Therefore, by Proposition 4.1,

$$\bar{x}^*_{(i,i+r]}(t_J^{\mathrm{cl}}) \geq \bar{x}_{(i,i+r]}(t_J^{\mathrm{cl}}) = \bar{x}_{(i+r,i+k]}(t_J^{\mathrm{cl}}) \geq \bar{x}^*_{(i+r,i+k]}(t_J^{\mathrm{cl}}).$$

Hence, $Q_{i,i+r,i+k}(t_J^{\mathrm{cl}}) \geq 0$, so that by (4.5), $t_J^{\mathrm{cl}} \geq \tau^*_{i,i+r,i+k}$. Since $r \in [1, k)$ is arbitrary, this yields $t_J^{\mathrm{cl}} \geq \max_{0<r<k} \tau^*_{i,i+r,i+k}$. $\square$

**5. Proofs.**

5.1. *Proof of Theorem* 3.3. In Theorem 3.3, we work with the cold gas, that is, there is no initial speed. Thus,

$$\tau^*_{\alpha,j,\beta} = \left(\frac{2n}{\gamma\rho(\beta-\alpha)}\right)^{1/2}(\bar{x}_{(j,\beta]}(0) - \bar{x}_{(\alpha,j]}(0))^{1/2}. \tag{5.1}$$

We will first provide a lower bound for $L_n(t)$ for the Poisson model, that is, if the initial positions $(x_i(0),\ 1\le i\le n)$ of the particles are the first $n$ jump times, denoted by $X_1 < \cdots < X_n$, of a Poisson process of intensity $n$. Recall that

$$L_n(t) = \max\Big\{k: \min_{J:|J|=k} t^{\mathrm{cl}}_J \le t\Big\}. \tag{5.2}$$

Let $k < n$. Consider the blocks $J_\ell := (\ell k, (\ell+1)k]$, $0 \le \ell \le \frac{n}{k}-1$. It follows from (5.2) that

$$\mathbb{P}(L_n(t) < k) \le \mathbb{P}\Big(\min_\ell t^{\mathrm{cl}}_{J_\ell} > t\Big). \tag{5.3}$$

For a generic block $J = (i, i+k]$, we have, by (5.1),

$$\begin{aligned}
b(J) &:= \max_{0<r<k} \tau^*_{i,i+r,i+k} \\
&= \left(\frac{2n}{\gamma\rho k}\right)^{1/2} \max_{0<r<k}\left[\frac{1}{k-r}\sum_{j=i+r+1}^{i+k}(X_j - X_i) - \frac{1}{r}\sum_{j=i+1}^{i+r}(X_j - X_i)\right]^{1/2}.
\end{aligned} \tag{5.4}$$

Taking $J = J_\ell = (\ell k, (\ell+1)k]$, and in view of (4.8), we have

$$\begin{aligned}
t^{\mathrm{cl}}_{J_\ell} \le b(J_\ell) &= \left(\frac{2n}{\gamma\rho k}\right)^{1/2} \max_{0<r<k}\bigg[\frac{1}{k-r}\sum_{j=\ell k+r+1}^{(\ell+1)k}(X_j - X_{\ell k}) \\
&\qquad\qquad - \frac{1}{r}\sum_{j=\ell k+1}^{\ell k+r}(X_j - X_{\ell k})\bigg]^{1/2},
\end{aligned} \tag{5.5}$$

with the notation $X_0 := 0$. Note that the random variables $b(J_\ell)$, for $0 \le \ell \le \frac{n}{k}-1$, are independent and identically distributed. Therefore, in light of (5.3), we obtain

$$\begin{aligned}
\mathbb{P}(L_n(t) < k) &\le \mathbb{P}\Big(\min_\ell b(J_\ell) > t\Big) \\
&= (1 - \mathbb{P}(b(J_0) \le t))^\nu \le \exp\{-\nu\mathbb{P}(b(J_0) \le t)\},
\end{aligned} \tag{5.6}$$

where $\nu := \lfloor \frac{n}{k} \rfloor$ denotes the number of blocks. According to (5.4),

$$\mathbb{P}(b(J_0) \leq t) = \mathbb{P}\left(\max_{0<r<k}\left(\frac{1}{k-r}\sum_{j=r+1}^{k} X_j - \frac{1}{r}\sum_{j=1}^{r} X_j\right) \leq \frac{\gamma\rho k}{2n}t^2\right).$$

At this stage, it is useful to recall from Fact 2.1 that

(5.7) $$X_j = \widetilde{X}_j X_{k+1}.$$

Since $(\widetilde{X}_1, \ldots, \widetilde{X}_k)$ and $X_{k+1}$ are independent, we arrive at the following estimate: for any $\delta > 0$,

(5.8)
$$\mathbb{P}(b(J_0) \leq t) \geq \mathbb{P}\left(\max_{0<r<k}\left(\frac{1}{k-r}\sum_{j=r+1}^{k} \widetilde{X}_j - \frac{1}{r}\sum_{j=1}^{r} \widetilde{X}_j\right) \leq \frac{1}{2} + \delta\right)$$

$$\times \mathbb{P}\left(X_{k+1} \leq \frac{\gamma\rho k}{(1+2\delta)n}t^2\right).$$

Consider the empirical quantile function

(5.9) $$\mathcal{Q}_k(s) := \inf\{t : F_k(t) \geq s\}, \qquad 0 \leq s \leq 1,$$

where $F_k$ is the empirical distribution function based on the random variables $(\widetilde{X}_j)_{1\leq j\leq k}$, that is, $F_k(t) := \frac{1}{k}\#\{j : 1 \leq j \leq k, \ \widetilde{X}_j \leq t\}$. Then

$$\widetilde{X}_j = \mathcal{Q}_k(s), \qquad \frac{j-1}{k} \leq s < \frac{j}{k},$$

and we have

$$\frac{1}{k-r}\sum_{j=r+1}^{k} \widetilde{X}_j - \frac{1}{r}\sum_{j=1}^{r}\widetilde{X}_j = \frac{1}{1-t}\int_t^1 \mathcal{Q}_k(s)\,ds - \frac{1}{t}\int_0^t \mathcal{Q}_k(s)\,ds, \qquad t := \frac{r}{k}.$$

Accordingly,

(5.10)
$$\max_{0<r<k}\left|\frac{1}{k-r}\sum_{j=r+1}^{k}\widetilde{X}_j - \frac{1}{r}\sum_{j=1}^{r}\widetilde{X}_j - \frac{1}{2}\right|$$

$$\leq \max_{0<t<1}\left|\frac{1}{1-t}\int_t^1 (\mathcal{Q}_k(s)-s)\,ds - \frac{1}{t}\int_0^t (\mathcal{Q}_k(s)-s)\,ds\right|.$$

The Glivenko–Cantelli theorem for quantile processes (see, e.g., [15], page 95) asserts that $\lim_{k\to\infty}\sup_s |\mathcal{Q}_k(s) - s| = 0$ almost surely. Hence,

$$\lim_{k\to\infty}\mathbb{P}\left(\max_{0<r<k}\left(\frac{1}{k-r}\sum_{j=r+1}^{k}\widetilde{X}_j - \frac{1}{r}\sum_{j=1}^{r}\widetilde{X}_j\right) \leq \frac{1}{2}+\delta\right) = 1.$$

For the second probability expression on the right-hand side of (5.8), we note that $\frac{\gamma\rho k}{(1+2\delta)n}t^2 = \frac{k}{(1+2\delta)n}(t/T^*)^2$. Therefore, for all sufficiently large $k$ (how large depending on $\delta$),

$$\mathbb{P}\left(X_{k+1} \leq \frac{\gamma\rho k}{(1+2\delta)n}t^2\right) = \mathbb{P}\left(\frac{nX_{k+1}}{k+1} \leq \frac{k}{(k+1)(1+2\delta)}(t/T^*)^2\right)$$

(5.11)

$$\geq \mathbb{P}\left(\frac{nX_{k+1}}{k+1} \leq (1+3\delta)^{-1}(t/T^*)^2\right).$$

By Chernoff's large deviation principle, for all large $k$,

$$\mathbb{P}\left(\frac{nX_{k+1}}{k+1} \leq (1+3\delta)^{-1}(t/T^*)^2\right) \geq \exp\{-(1+\delta)I((1+3\delta)^{-1}(t/T^*)^2)(k+1)\},$$

where the large deviation rate function $I(\cdot)$ of the exponential law is as in (3.3). We choose now

$$k = k(n) \sim \frac{\log n}{(1+3\delta)I((1+3\delta)^{-1}(t/T^*)^2)},$$

so that by (5.8),

$$\nu\mathbb{P}(b(J_0) \leq t) \geq \text{const} \times \frac{n}{\log n} n^{-(1+\delta)/(1+2\delta)} \to \infty.$$

We derive via (5.6) that $\mathbb{P}(L_n(t) < k) \to 0$. Therefore,

$$\liminf_{n\to\infty} \frac{L_n(t)}{\log n} \geq (1+3\delta)^{-1}I((1+3\delta)^{-1}(t/T^*)^2)^{-1} \qquad \text{in probability.}$$

By sending $\delta$ to zero, we obtain the desired lower bound in Theorem 3.3 for the Poisson model:

$$\liminf_{n\to\infty} \frac{L_n(t)}{\log n} \geq I((t/T^*)^2)^{-1} \qquad \text{in probability.}$$

We prove now the upper bound for the Poisson model. Let $k < n$. Assume that we are in the situation $L_n(t) > k$, which means that prior to (or at) time $t$, a cluster (say, $A$) of size $k+1$ or larger appears. Actually, one can choose $A$ of size less than or equal to $2k$, but we do not need the upper bound in *this* proof. Consider the block of particles $J$ which corresponds to $A$ in a natural way. Notice that $J$ is free up to the time of formation of $A$. By (5.4) and (4.9), we obtain

(5.12) $$b(J) = t_J^{\text{cl}} \leq t.$$

To bound $b(J)$ from below, we use the following elementary estimate.

LEMMA 5.1. *Let $p \geq 2$ be an integer, and let $x_1 \leq \cdots \leq x_p$ be real numbers. Then*

$$\max_{0<r<p}\left(\frac{1}{p-r}\sum_{j=r+1}^{p} x_j - \frac{1}{r}\sum_{j=1}^{r} x_j\right) \geq \frac{x_p - x_1}{2}.$$

PROOF. For $p = 2$, the left-hand side just equals $x_p - x_1$. For $p > 2$, consider the average $\bar{x} := \frac{1}{p-2}\sum_{j=2}^{p-1} x_j$. Assume first that $\bar{x} \leq \frac{x_1+x_p}{2}$. Then

$$\max_{0<r<p}\left(\frac{1}{p-r}\sum_{j=r+1}^{p} x_j - \frac{1}{r}\sum_{j=1}^{r} x_j\right) \geq x_p - \frac{1}{p-1}\sum_{j=1}^{p-1} x_j$$

$$= x_p - \frac{1}{p-1}((p-2)\bar{x} + x_1)$$

$$\geq x_p - \frac{p-2}{2(p-1)}(x_1 + x_p) - \frac{x_1}{p-1}$$

$$= \frac{p}{2(p-1)}(x_p - x_1) \geq \frac{x_p - x_1}{2}.$$

The case $\bar{x} \geq \frac{x_1+x_p}{2}$ boils down to the considered one by the substitution of $x_j$ by $-x_j$. □

The definition of $b(J)$ and Lemma 5.1 yield for any block $J = (i, i+p]$,

$$b(J) \geq \left(\frac{2n}{\gamma\rho p}\right)^{1/2}\left(\frac{X_{i+p} - X_{i+1}}{2}\right)^{1/2}.$$

Plugging this into (5.12) yields that for any $k$ and $t$,

(5.13) $\{L_n(t) > k\} \subset \bigcup_{p=k+1}^{n}\bigcup_{i=0}^{n-p}\left\{\left(\frac{2n}{\gamma\rho p}\right)^{1/2}\left(\frac{X_{i+p} - X_{i+1}}{2}\right)^{1/2} \leq t\right\}.$

Since the random variables $X_{i+p} - X_{i+1}$, $0 \leq i \leq n-p$, are identically distributed, we get

$$\mathbb{P}(L_n(t) > k) \leq n\sum_{p=k+1}^{n}\mathbb{P}\left(\left(\frac{2n}{\gamma\rho p}\right)^{1/2}\left(\frac{X_p - X_1}{2}\right)^{1/2} \leq t\right)$$

$$= n\sum_{p=k+1}^{n}\mathbb{P}\left(X_p - X_1 \leq \frac{\gamma\rho p}{n}t^2\right).$$

The random variable $n(X_p - X_1)$ is distributed as the sum of $p - 1$ independent random variables having the standard exponential distribution.

Therefore, Chernoff's large deviation principle yields that for any fixed $\delta > 0$ and all sufficiently large $k$ and all $p > k$,

$$\mathbb{P}\left(X_p - X_1 \leq \frac{\gamma \rho p}{n} t^2\right) = \mathbb{P}\left(X_p - X_1 \leq \frac{p}{n}(t/T^*)^2\right)$$
$$\leq \mathbb{P}\left(\frac{n(X_p - X_1)}{p-1} \leq (1+\delta)(t/T^*)^2\right)$$
$$\leq \exp\{-(1-\delta)I((1+\delta)(t/T^*)^2)(p-1)\},$$

from which it follows that

$$\mathbb{P}(L_n(t) > k) \leq n \sum_{p=k+1}^{\infty} \exp\{-(1-\delta)I((1+\delta)(t/T^*)^2)(p-1)\}$$
$$= \frac{n \exp\{-(1-\delta)I((1+\delta)(t/T^*)^2)k\}}{1 - \exp\{-(1-\delta)I((1+\delta)(t/T^*)^2)\}}.$$

We choose now

$$k = k(n) \sim \frac{(1+\delta)\log n}{(1-\delta)I((1+\delta)(t/T^*)^2)}, \qquad n \to \infty,$$

so that $\mathbb{P}(L_n(t) > k) \to 0$, $n \to \infty$. Therefore,

$$\limsup_{n \to \infty} \frac{L_n(t)}{\log n} \leq \frac{1+\delta}{1-\delta} I((1+\delta)(t/T^*)^2)^{-1} \qquad \text{in probability.}$$

Finally, by sending $\delta$ to zero, we obtain the desired estimate

$$\limsup_{n \to \infty} \frac{L_n(t)}{\log n} \leq I((t/T^*)^2)^{-1} \qquad \text{in probability.}$$

This completes the proof of Theorem 3.3 for the Poisson model. In order to prove it for the i.i.d.-model, we use the representation from Fact 2.1 with a Poisson process of intensity $n$. The system with initial positions $(\widetilde{X}_i)_{i \leq n}$ is similar to the Poisson model with initial positions $(X_i)_{i \leq n}$, if we take similarity coefficients as follows: $c_x = (X_{n+1})^{-1}$, $c_v = (X_{n+1})^{-1/2}$, $c_t = (X_{n+1})^{-1/2}$. We infer from the similarity that

(5.14) $$\widetilde{L}_n(t) = L_n(t\sqrt{X_{n+1}}),$$

where $\widetilde{L}_n$ and $L_n$ denote the size of the maximal cluster in the i.i.d.-model and in the Poisson model, respectively. Taking into account that $X_{n+1} \to 1$ in probability (law of large numbers), we derive the statement of the theorem for the i.i.d.-model from what we have just proved for the Poisson model. □

5.2. *Proof of Theorem* 3.4. Consider first the Poisson model. Let $\varepsilon_n := \frac{T^* - t_n}{T^*}$. The assumptions in the theorem take the form

(5.15) $$\lim_{n \to \infty} \varepsilon_n = 0,$$

(5.16) $$\lim_{n \to \infty} n\varepsilon_n^2 = \infty.$$

We fix a small $\beta \in (0, 2)$ and take a sequence of positive integers $(k_n)$ such that

$$k_n \sim \frac{\log(n\varepsilon_n^2)}{(2-\beta)\varepsilon_n^2}, \qquad n \to \infty.$$

It follows from (5.13) that

$$\{L_n(t_n) > k_n\} \subset \bigcup_{p=k_n+1}^{n} \bigcup_{i=0}^{n-p} \left\{ n(X_{i+p} - X_{i+1}) \leq \frac{t_n^2 p}{(T^*)^2} \right\}.$$

It is easier to interpret the condition within the brackets on the right-hand side in terms of centered random variables, namely,

$$(nX_{i+p} - (i+p)) - (nX_{i+1} - (i+1)) \leq \frac{t_n^2 p}{(T^*)^2} - p + 1 = -p\varepsilon_n(2 - \varepsilon_n) + 1.$$

Recall that for any $n \geq 1$, the random variables $(nX_i - i)_{i \geq 0}$ can be viewed as sums of independent centered random variables having the standard centered exponential distribution. Using the classical Komlós–Major–Tusnády estimate [8], we approximate them by a Wiener process $W$. Namely, we need the following:

FACT 5.2 (KMT construction). It is possible to construct $W$ and $(X_i)$ in a common probability space such that for some numerical constants $C_1$ and $C_2$, and for all real $r > 0$,

(5.17) $$\mathbb{P}\left\{ \max_{1 \leq i \leq n} |(nX_i - i) - W(i)| > r \right\} \leq (1 + C_1 n^{1/2}) \exp\{-C_2 r\}.$$

Since we study only convergence in probability, it is unimportant for us whether the probability space in KMT construction is the same for all $n$ or not. We just make use of

$$\lim_{n \to \infty} \mathbb{P}\left\{ \max_{1 \leq i \leq n} |(nX_i - i) - W(i)| > \frac{\log n}{C_2} \right\} = 0.$$

Observe that

$$\mathbb{P}\{L_n(t_n) > k_n\}$$

$$\leq \mathbb{P}\left\{\bigcup_{p=k_n+1}^{n}\bigcup_{i=0}^{n-p}\left\{|W(i+p)-W(i)|\geq p\varepsilon_n(2-\varepsilon_n)-1-\frac{2}{C_2}\log n\right\}\right\}$$

$$+\mathbb{P}\left\{\max_{1\leq i\leq n}|(nX_i-i)-W(i)|>\frac{\log n}{C_2}\right\}.$$

The second term on the right-hand side converges to zero by what we have just seen. On the other hand, under assumptions (5.15)–(5.16), we have $p\varepsilon_n \geq k_n\varepsilon_n \sim \frac{\log(n\varepsilon_n^2)}{(2-\beta)\varepsilon_n} \gg \log n$. Therefore, for any fixed $h > 0$,

$$\limsup_{n\to\infty}\mathbb{P}\{L_n(t_n)>k_n\}$$

$$\leq \limsup_{n\to\infty}\mathbb{P}\left\{\bigcup_{p=k_n+1}^{n}\bigcup_{i=0}^{n-p}\{|W(i+p)-W(i)|\geq p\varepsilon_n(2-h)\}\right\}.$$

By scaling, the probability expression on the right-hand side is

$$=\mathbb{P}\left\{\bigcup_{p=k_n+1}^{n}\bigcup_{i=0}^{n-p}\left\{\left(\frac{p}{k_n}\right)^{-1/2}\left|W\left(\frac{i}{k_n}+\frac{p}{k_n}\right)-W\left(\frac{i}{k_n}\right)\right|\geq\sqrt{p}\varepsilon_n(2-h)\right\}\right\}$$

$$\leq \mathbb{P}\left\{\sup_{0\leq t\leq n/k_n}\sup_{1\leq u\leq n/k_n}u^{-1/2}|W(t+u)-W(t)|\geq\sqrt{k_n}\varepsilon_n(2-h)\right\}.$$

We write, for any $a > 0$ and $b \geq 1$,

(5.18) $$\Delta(a,b) := \sup_{0\leq t\leq a}\sup_{1\leq u\leq b}u^{-1/2}|W(t+u)-W(t)|.$$

Then

(5.19) $$\limsup_{n\to\infty}\mathbb{P}\{L_n(t_n)>k_n\}\leq\limsup_{n\to\infty}\mathbb{P}\left\{\Delta\left(\frac{n}{k_n},\frac{n}{k_n}\right)\geq\sqrt{k_n}\varepsilon_n(2-h)\right\}.$$

The following lemma gives an upper bound for the tail of $\Delta$.

LEMMA 5.3. *Let $\Delta(\cdot,\cdot)$ be as in (5.18). For any $h \in (0,1)$, there exists $c_h > 0$ such that for all $T \geq 1$, $U \geq 2$ and $r \geq 1$,*

$$\mathbb{P}(\Delta(T,U)\geq r)\leq c_h T(\log U)\exp\{-(1-h)r^2/2\}.$$

PROOF. By scaling, for any $a \geq 1$,

$$\mathbb{P}\left(\sup_{0\leq s\leq T}\sup_{a\leq u\leq 2a}\frac{|W(s+u)-W(s)|}{u^{1/2}}\geq r\right)=\mathbb{P}(\Delta(T/a,2)\geq r)$$

$$\leq \mathbb{P}(\Delta(T,2)\geq r).$$

Therefore, by the stationarity of the increments of the Wiener process,

(5.20) $$\mathbb{P}(\Delta(T,U)\geq r)\leq \lceil T\rceil\left\lceil\frac{\log U}{\log 2}\right\rceil\mathbb{P}(\Delta(1,2)\geq r).$$

On the other hand, for any centered Gaussian process $\{Y(v), v \in V\}$, the following estimate of large deviations holds (see [9], Chapter 12):

$$\lim_{r \to \infty} r^{-2} \log \mathbb{P}\left(\sup_{v \in V} |Y(v)| \geq r\right) = -\frac{1}{2\sigma^2},$$

where $\sigma^2 := \sup_{v \in V} \mathbb{E}[Y(v)^2]$. We apply this estimate to the two-parameter process

$$Y(s, u) := \frac{W(s+u) - W(s)}{u^{1/2}}, \qquad (s, u) \in V := [0,1] \times [1,2],$$

so that $\lim_{r \to \infty} r^{-2} \log \mathbb{P}(\Delta(1,2) \geq r) = -\frac{1}{2}$. Therefore,

$$\sup_{r \geq 1} \frac{\mathbb{P}(\Delta(1,2) \geq r)}{\exp\{-(1-h)r^2/2\}} < \infty.$$

This, in light of (5.20), yields the lemma. $\square$

Let now $h$ be so small that

$$z := \frac{(1-h)(2-h)^2}{2(2-\beta)} > 1.$$

Applying Lemma 5.3 to $T = U = n/k_n$ and $r = \sqrt{k_n}(2-h)\varepsilon_n$, we get

$$\mathbb{P}\left\{\Delta\left(\frac{n}{k_n}, \frac{n}{k_n}\right) \geq \sqrt{k_n}\varepsilon_n(2-h)\right\}$$

$$\leq c_h \frac{n \log(n/k_n)}{k_n} \exp\{-(1-h)(2-h)^2 \varepsilon_n^2 k_n / 2\}$$

$$= c_h \frac{n}{k_n} (n\varepsilon_n^2)^{-(1+o(1))z} \log(n/k_n)$$

$$\sim c_h (2-\beta)(n\varepsilon_n^2)^{1-(1+o(1))z}.$$

(We have used the fact that $n\varepsilon_n^2 \to \infty$.) Since $z > 1$, we have $(n\varepsilon_n^2)^{1-(1+o(1))z} \to 0$. Plugging this into (5.19) gives $\lim_{n \to \infty} \mathbb{P}\{L_n(t_n) \geq k_n\} = 0$, which means that

$$1 \geq \limsup_{n \to \infty} \frac{L_n(t_n)}{k_n}$$

$$= (2-\beta) \limsup_{n \to \infty} \frac{L_n(t_n)\varepsilon_n^2}{\log(n\varepsilon_n^2)}$$

$$= \frac{2-\beta}{(T^*)^2} \limsup_{n \to \infty} \frac{L_n(t_n)(T^* - t_n)^2}{\log(n(T^* - t_n)^2)} \qquad \text{in probability.}$$

The desired upper bound for the Poisson model follows by letting $\beta \to 0$.

We proceed now to the proof of the lower bound for the Poisson model. We have already established in inequalities (5.6), (5.8) and (5.11) that for all $t > 0$, $\delta > 0$ and all integer $k$ (such that $\frac{k}{k+1} \geq \frac{1+2\delta}{1+3\delta}$),

(5.21) $$\mathbb{P}(L_n(t) < k) \leq \exp\{-\nu P_1(k, \delta) P_2(k, \delta, t)\},$$

where $\nu := \lfloor \frac{n}{k} \rfloor$ and

$$P_1(k, \delta) := \mathbb{P}\left(\max_{0 < r < k}\left(\frac{1}{k-r}\sum_{j=r+1}^{k}\widetilde{X}_j - \frac{1}{r}\sum_{j=1}^{r}\widetilde{X}_j\right) \leq \frac{1}{2} + \delta\right),$$

$$P_2(k, \delta, t) := \mathbb{P}\left(\frac{nX_{k+1}}{k+1} \leq (1+3\delta)^{-1}(t/T^*)^2\right).$$

Let us now specify the choice of the parameters. Let $\beta > 0$ and $h > 0$ be some small numbers. We still denote $\varepsilon_n := \frac{T^* - t_n}{T^*}$, but this time we set $(k_n)$ such that

$$k_n \sim \frac{\log(n\varepsilon_n^2)}{(2+\beta)\varepsilon_n^2}.$$

Let $\delta_n := h\varepsilon_n$. Then

$$P_2(k_n, \delta_n, t_n) = \mathbb{P}\left(\frac{nX_{k_n+1} - (k_n+1)}{\sqrt{k_n+1}} \leq ((1+3\delta_n)^{-1}(t_n/T^*)^2 - 1)\sqrt{k_n+1}\right).$$

For all large $n$,

$$((1+3\delta_n)^{-1}(t_n/T^*)^2 - 1)\sqrt{k_n+1}$$
$$\geq ((1-3\delta_n)(1-\varepsilon_n)^2 - 1)\sqrt{k_n+1}$$
$$\geq -(3\delta_n + 2\varepsilon_n)\sqrt{k_n+1}$$
$$= -(3h+2)\varepsilon_n\sqrt{k_n+1}.$$

Therefore, for all large $n$,

$$P_2(k_n, \delta_n, t_n) \geq \mathbb{P}\left(\frac{nX_{k_n+1} - (k_n+1)}{\sqrt{k_n+1}} \leq -(3h+2)\varepsilon_n\sqrt{k_n+1}\right).$$

Now we use the following result of moderate deviations (see [12], Chapter 8, page 218).

FACT 5.4 (Cramér's theorem). Let $Y_i$ be a sequence of i.i.d. centered random variables such that $\mathbb{E}\exp\{\gamma|Y_i|\} < \infty$ for some $\gamma > 0$. Then for any positive sequence $(x_n)$ with $x_n = o(\sqrt{n})$, we have

$$\mathbb{P}\left\{\frac{1}{\sqrt{n}}\sum_{j=1}^{n}Y_j > x_n\right\} = \exp\left(-\frac{1+o(1)}{2}x_n^2\right).$$

Since $\varepsilon_n \to 0$ [see (5.15)], we are entitled to apply Fact 5.4 to $x_n = (3h+2)\varepsilon_n\sqrt{k_n+1}$. Note that $\varepsilon_n\sqrt{k_n+1} \to \infty$ [see (5.16)]. Thus,

$$P_2(k_n,\delta_n,t_n) \geq (1+o(1))\exp\{-(3h+2)^2\varepsilon_n^2 k_n(1+o(1))/2\}$$
$$\geq (1+o(1))\exp\left\{-\frac{(3h+2)^2}{2(2+\beta)}\log(n\varepsilon_n^2)(1+o(1))\right\}$$
$$= (1+o(1))(n\varepsilon_n^2)^{-(1+o(1))z},$$

where $z := \frac{(3h+2)^2}{2(2+\beta)}$. The parameter $h$ can be chosen so small that $z < 1$. With this choice, (5.16) implies

(5.22) $$\nu P_2(k_n,\delta_n,t_n) \geq \frac{n}{2k_n}(n\varepsilon_n^2)^{-(1+o(1))z} \sim \frac{(2+\beta)(n\varepsilon_n^2)^{1-(1+o(1))z}}{2\log(n\varepsilon_n^2)} \to \infty.$$

Assume for a while that

(5.23) $$\liminf_{n\to\infty} P_1(k_n,\delta_n) > 0.$$

Then by (5.21) and (5.22), $\lim_{n\to\infty}\mathbb{P}\{L_n(t_n) \leq k_n\} = 0$, which implies the desired lower bound:

$$\liminf_{n\to\infty}\frac{L_n(t_n)}{k_n} \geq 1 \quad \text{in probability.}$$

To establish (5.23), we observe that

(5.24) $$\delta_n k_n^{1/2} = h\varepsilon_n\sqrt{k_n} \sim h\left(\frac{\log(n\varepsilon_n^2)}{2+\beta}\right)^{1/2} \to \infty.$$

We use the estimate (5.10), where $\mathcal{Q}_k(\cdot)$ is the empirical quantile function defined in (5.9). The well-known functional limit theorem for quantile processes (see [15]) asserts that for $k \to \infty$, the sequence of processes $Y_k(r) = \sqrt{k}(\mathcal{Q}_k(r) - r)$, $r \in [0,1]$, converges weakly to a Brownian bridge $\overset{\circ}{W}$ in the Skorokhod topology. In light of (5.24), we arrive at

$$\liminf_{n\to\infty} P_1(k_n,\delta_n)$$
$$\geq \liminf_{n\to\infty}\mathbb{P}\left(\max_{0<t<1}\left|\frac{1}{1-t}\int_t^1(\mathcal{Q}_{k_n}(s)-s)\,ds - \frac{1}{t}\int_0^t(\mathcal{Q}_{k_n}(s)-s)\,ds\right| \leq \delta_n\right)$$
$$\geq \liminf_{n\to\infty}\mathbb{P}\left(\max_{0<t<1}\left|\frac{1}{1-t}\int_t^1 Y_{k_n}(s)\,ds - \frac{1}{t}\int_0^t Y_{k_n}(s)\,ds\right| < 1\right)$$
$$= \mathbb{P}\left(\max_{0<t<1}\left|\frac{1}{1-t}\int_t^1 \overset{\circ}{W}(s)\,ds - \frac{1}{t}\int_0^t \overset{\circ}{W}(s)\,ds\right| < 1\right) > 0.$$

Therefore, relation (5.23) is true and Theorem 3.4 is proved for the Poisson model.

Consider now the i.i.d.-model. We use (5.14) which reduces the problem to the Poisson case. Indeed, by the central limit theorem, for any $\varepsilon > 0$, there exists $M > 0$ such that

$$\limsup_{n \to \infty} \mathbb{P}\left(|\sqrt{X_{n+1}} - 1| > \frac{M}{\sqrt{n}}\right) < \varepsilon.$$

Let $\hat{t}_n := t_n(1 + \frac{M}{\sqrt{n}})$. The sequence $(\hat{t}_n)$ satisfies the assumptions in Theorem 3.4, and the norming sequences are equivalent:

$$\lim_{n \to \infty} \frac{(T^* - t_n)^{-2} \log(n(T^* - t_n)^2)}{(T^* - \hat{t}_n)^{-2} \log(n(T^* - \hat{t}_n)^2)} = 1.$$

By (5.14), if $\sqrt{X_{n+1}} \le 1 + \frac{M}{\sqrt{n}}$, then $L_n(\hat{t}_n) \ge L_n(t_n\sqrt{X_{n+1}}) = \widetilde{L}_n(t_n)$. Therefore, for any $\beta > 0$, we have, for all large $n$,

$$\mathbb{P}\left(\widetilde{L}_n(t_n) > \frac{(T^*)^2 \log(n(T^* - t_n)^2)}{(2 + \beta)(T^* - t_n)^2}\right)$$

$$\le \mathbb{P}\left(\sqrt{X_{n+1}} > 1 + \frac{M}{\sqrt{n}}\right) + \mathbb{P}\left(L_n(\hat{t}_n) > \frac{(T^*)^2 \log(n(T^* - \hat{t}_n)^2)}{(2 + 2\beta)(T^* - \hat{t}_n)^2}\right).$$

The last probability term on the right-hand side tends to zero by what we have proved for the Poisson case. Thus,

$$\limsup_{n \to \infty} \mathbb{P}\left(\widetilde{L}_n(t_n) > \frac{(T^*)^2 \log(n(T^* - t_n)^2)}{(2 + \beta)(T^* - t_n)^2}\right)$$

$$\le \limsup_{n \to \infty} \mathbb{P}\left(\sqrt{X_{n+1}} > 1 + \frac{M}{\sqrt{n}}\right) \le \varepsilon.$$

Since $\varepsilon$ is arbitrary, we get the desired upper bound: for any $\beta > 0$,

$$\limsup_{n \to \infty} \mathbb{P}\left(\widetilde{L}_n(t_n) > \frac{(T^*)^2 \log(n(T^* - t_n)^2)}{(2 + \beta)(T^* - t_n)^2}\right) = 0.$$

The lower bound follows exactly in the same way. $\square$

5.3. *Proof of Theorem* 3.5. We first focus on the lattice model. In this model, for every block $J = (i, i + k]$, we have

(5.25) $$\bar{x}_J(0) = \frac{1}{k} \sum_{j=i+1}^{i+k} x_j(0) = \frac{1}{k} \sum_{j=i+1}^{i+k} \frac{j}{n} = \frac{2i + k + 1}{2n}.$$

Let

$$\bar{u}_J := \frac{1}{M_J} \sum_{j \in J} m_j u_j = \frac{1}{k} \sum_{j=i+1}^{i+k} u_j,$$

AGGREGATION IN SELF-GRAVITATING GAS 23which denotes the average of normalized initial speeds. Then (4.1) takes the form

$$\bar{x}_J^*(s) = \frac{2i+k+1}{2n} + \sigma_n \bar{u}_J s + \gamma\rho(n-2i-k)\frac{s^2}{2n}, \tag{5.26}$$

and (4.5) becomes, for any $t > 0$,

$$\tau_{\alpha,j,\beta}^* \leq t \iff \bar{u}_{(\alpha,j]} - \bar{u}_{(j,\beta]} \geq \frac{(\beta-\alpha)(1-\gamma\rho t^2)}{2n\sigma_n t}. \tag{5.27}$$

We start with the proof of the lower bound for $L_n(t)$. Let $\varepsilon \in (0, 1/3)$ and let $k$ be an arbitrary positive integer. Consider $J_\ell = [(\ell+\varepsilon)k, (\ell+1-\varepsilon)k] \cap \mathbb{Z}$. According to (4.7), for any $\ell$, if there exist $\alpha < (\ell+\varepsilon)k$ and $\beta > (\ell+1-\varepsilon)k$ such that $\max_{j \in J_\ell} \tau_{\alpha,j,\beta}^* \leq t$, then all the particles of the block $J_\ell$ belong to a common cluster at time $t$. We apply this to $\alpha = \ell k$ and $\beta = (\ell+1)k$ to see, in light of (5.27), that the condition

$$\min_{j \in J_\ell}(\bar{u}_{(\ell k, j]} - \bar{u}_{(j,(\ell+1)k]}) \geq \frac{k(1-\gamma\rho t^2)}{2n\sigma_n t}$$

ensures a single cluster at time $t$ from the block $J_\ell$. Thus, similarly to (5.6), we have

$$\mathbb{P}(L_n(t) \leq (1-2\varepsilon)k - 1) \leq \exp\{-\nu P_1\}, \tag{5.28}$$

where $\nu := \lfloor \frac{n}{k} \rfloor$ denotes the number of blocks and

$$P_1 := \mathbb{P}\bigg(\min_{j \in J_0}(\bar{u}_{(0,j]} - \bar{u}_{(j,k]}) \geq \frac{k(1-\gamma\rho t^2)}{2n\sigma_n t}\bigg).$$

Let us express the average speeds in terms of the random walk

$$S_j := \sum_{i=1}^{j} u_i, \qquad j \geq 0.$$

Then $\bar{u}_{(0,j]} - \bar{u}_{(j,k]} = \frac{kS_j - jS_k}{j(k-j)}$ and, thus,

$$P_1 = \mathbb{P}\bigg\{k^{-1/2}\bigg(S_j - \frac{j}{k}S_k\bigg) \geq \frac{k^{3/2}(1-\gamma\rho t^2)}{2n\sigma_n t}\frac{j}{k}\bigg(1-\frac{j}{k}\bigg), \\ \varepsilon \leq \frac{j}{k} \leq 1-\varepsilon\bigg\}. \tag{5.29}$$

Let us now specify the choice of $k$ by letting

$$b = b_n := \log\bigg(\frac{n}{\sigma_n^2}\bigg),$$

$$k = k_n \sim (1-4\varepsilon)^{2/3} c_1(t)(n\sigma_n)^{2/3} b^{1/3},$$

where $c_1(t)$ is defined in (3.8). Note that by the definition of $c_1(t)$, $\frac{k^{3/2}(1-\gamma\rho t^2)}{2n\sigma_n t} \sim (1-4\varepsilon)(2b)^{1/2}$. Accordingly, for all large $n$,

$$P_1 \geq \mathbb{P}\left\{k^{-1/2}\left(S_j - \frac{j}{k}S_k\right) \geq (1-3\varepsilon)(2b)^{1/2}\frac{j}{k}\left(1-\frac{j}{k}\right),\ \varepsilon \leq \frac{j}{k} \leq 1-\varepsilon\right\}.$$

Let $W$ be the Wiener process in (5.17). We have

$$P_1 \geq \mathbb{P}\Big\{k^{-1/2}\left(W(j) - \frac{j}{k}W(k)\right) \geq (1-\varepsilon)(2b)^{1/2}\frac{j}{k}\left(1-\frac{j}{k}\right),$$
$$k^{-1/2}|W(j) - S_j| \leq \varepsilon(2b)^{1/2}\frac{j}{k}\left(1-\frac{j}{k}\right),\ \varepsilon \leq \frac{j}{k} \leq 1-\varepsilon\Big\}$$
$$\geq \mathbb{P}\Big\{k^{-1/2}(W(ks) - sW(k)) \geq (1-\varepsilon)(2b)^{1/2}s(1-s),\ \varepsilon \leq s \leq 1-\varepsilon,$$
$$k^{-1/2}\max_{1\leq j\leq k}|W(j) - S_j| \leq \varepsilon^2(1-\varepsilon)(2b)^{1/2}\Big\}.$$

Since $s \mapsto k^{-1/2}W(ks)$, $s \geq 0$, is also a Wiener process, we get $P_1 \geq P_2 - P_3$, where

$$P_2 := \mathbb{P}\{W(s) - sW(1) \geq (1-\varepsilon)(2b)^{1/2}s(1-s),\ \varepsilon \leq s \leq 1-\varepsilon\},$$
$$P_3 := \mathbb{P}\left\{\max_{1\leq j\leq k}|W(j) - S_j| \geq \varepsilon^2(1-\varepsilon)(2bk)^{1/2}\right\}.$$

In order to estimate $P_2$, we apply the functional large deviation principle (see, e.g., [9], Chapter 12, Theorem 6) to the Brownian bridge $\overset{\circ}{W}(s) = W(s) - sW(1)$:

$$\liminf_{R\to\infty} \frac{1}{R^2}\log\mathbb{P}\{W(s) - sW(1) \geq Rs(1-s),\ \varepsilon \leq s \leq 1-\varepsilon\}$$
$$\geq -\frac{1}{2}\int_0^1 (1-2s)^2\,ds = -\frac{1}{6}.$$

By (3.5), we have $b \to \infty$, which implies that for large $n$,

(5.30) $$P_2 \geq \exp\{-(1+\varepsilon)^2(1-\varepsilon)^2 b/3\} = \left(\frac{n}{\sigma_n^2}\right)^{-(1-\varepsilon^2)^2/3}.$$

We now show that $P_3$ is negligible compared to $P_2$. Indeed, by the Komlós–Major–Tusnády estimate (5.17),

(5.31) $$P_3 \leq (1 + C_1 k^{1/2})\exp\{-C_2\varepsilon^2(1-\varepsilon)(2bk)^{1/2}\}.$$

Since $\frac{k}{b} \sim (1-4\varepsilon)^{2/3}c_1(t)(\frac{n\sigma_n}{b_n})^{2/3} \gg 1$, we have, for all sufficiently large $n$, $C_2\varepsilon^2(1-\varepsilon)(2bk)^{1/2} > b + \log(1 + C_1 k^{1/2})$, which implies that for all large $n$,

$$P_3 \leq \exp\{-b\} = \left(\frac{n}{\sigma_n^2}\right)^{-1}.$$

In view of (5.30), this yields $P_2 \gg P_3$. Hence,

$$\nu P_1 \geq \nu(P_2 - P_3)$$

$$\sim \frac{n}{k} P_2 \geq \frac{n}{c_1(t)(n\sigma_n)^{2/3}(\log(n/\sigma_n^2))^{1/3}} \left(\frac{n}{\sigma_n^2}\right)^{-(1-\varepsilon^2)^2/3}.$$

The expression on the right-hand side equals $\frac{1}{c_1(t)}(\frac{n}{\sigma_n^2})^{[1-(1-\varepsilon^2)^2]/3}(\log(\frac{n}{\sigma_n^2}))^{-1/3}$, and thus goes to infinity. In light of (5.28), we obtain

$$\liminf_{n\to\infty} \frac{L_n(t)}{(n\sigma_n)^{2/3}(\log(n/\sigma_n^2))^{1/3}} \geq (1-2\varepsilon)(1-4\varepsilon)^{2/3}c_1(t) \quad \text{in probability.}$$

The lower bound in Theorem 3.5 for the lattice model is proved, since $\varepsilon$ can be arbitrarily small.

We proceed now to the proof of the upper bound (for the lattice model). We have already seen in the proof of Theorem 3.3 that if $L_n(t) > k$, then there exists a block $J = (i, i+p]$ of size $p$ between $k+1$ and $2k$, that is free up to its collapse time $t_J^{\text{cl}}$ and such that $t_J^{\text{cl}} \leq t$. Let, for simplicity, $p$ be even, say, $p = 2q$. By virtue of Proposition 4.4, inequality $t_J^{\text{cl}} \leq t$ means that

$$\max_{0<r<2q} \tau^*_{i,i+r,i+2q} \leq t.$$

In particular, we have $\tau^*_{i,i+q,i+2q} \leq t$, which, in light of (5.27), is equivalent to

$$\bar{u}_{(i,i+q]} - \bar{u}_{(i+q,i+2q]} \geq \frac{q(1-\gamma\rho t^2)}{n\sigma_n t}.$$

Similarly, whenever $p$ is odd, say, $p = 2q+1$, we arrive at

$$\bar{u}_{(i,i+q]} - \bar{u}_{(i+q+1,i+2q+1]} \geq \frac{q(1-\gamma\rho t^2)}{n\sigma_n t}.$$

Summarizing, we have

$$\{L_n(t) > k\} \subset \Omega_1 \cup \Omega_2,$$

where

$$\Omega_1 = \bigcup_{q=k/2}^{k} \bigcup_{i=0}^{n-2q} \left\{ \frac{1}{\sqrt{2q}} \sum_{j=1}^{q} (u_{i+j} - u_{i+q+j}) \geq \frac{q^{3/2}(1-\gamma\rho t^2)}{\sqrt{2}n\sigma_n t} \right\}$$

and

$$\Omega_2 = \bigcup_{q=k/2}^{k} \bigcup_{i=0}^{n-2q} \left\{ \frac{1}{\sqrt{2q}} \sum_{j=1}^{q} (u_{i+j} - u_{i+q+1+j}) \geq \frac{q^{3/2}(1-\gamma\rho t^2)}{\sqrt{2}n\sigma_n t} \right\}.$$

Thus,
$$\mathbb{P}\{L_n(t) > k\} \leq 2n \sum_{q=k/2}^{k} P_q,$$

where
$$P_q := \mathbb{P}\left(\frac{1}{\sqrt{2q}} \sum_{j=1}^{q}(u_j - u_{q+j}) \geq x_{q,n}\right)$$

and
$$x_{q,n} = \frac{q^{3/2}(1 - \gamma\rho t^2)}{\sqrt{2}n\sigma_n t}.$$

Let $\varepsilon > 0$, and let
$$k = k(n) \sim (1 + 2\varepsilon)c_2(t)(n\sigma_n)^{2/3}(\log(n\sigma_n^{2/5}))^{1/3},$$

where $c_2(t)$ is defined in (3.8). Since we want to apply Fact 5.4 to $P_q$, let us check that

(5.32) $$\lim_{n\to\infty} \frac{x_{q,n}}{q^{1/2}} = 0$$

uniformly over $q \leq 2k(n)$. Indeed,
$$\frac{x_{q,n}}{q^{1/2}} = \mathcal{O}\left(\frac{q}{n\sigma_n}\right) = \mathcal{O}\left(\left(\frac{\log(n\sigma_n^{2/5})}{n\sigma_n}\right)^{1/3}\right).$$

By using twice the assumption (3.5), we have
$$\limsup_{n\to\infty} \frac{\log(n\sigma_n^{2/5})}{n\sigma_n} \leq \limsup_{n\to\infty} \frac{2\log(n)}{n\sigma_n} = 0$$

and (5.32) follows. Now, by Fact 5.4, for all large $n$ and all $q \in [k/2, k]$,
$$P_q \leq \exp\left\{\frac{-x_{q,n}^2}{2(1+\varepsilon)}\right\} = \exp\left\{-\frac{1}{2(1+\varepsilon)}\frac{q^3(1-\gamma\rho t^2)^2}{2(n\sigma_n)^2 t^2}\right\},$$

so that
$$\mathbb{P}\{L_n(t) > k\} \leq 2n \sum_{q \geq k/2} \exp\left\{-\frac{1}{2(1+\varepsilon)}\frac{q^3(1-\gamma\rho t^2)^2}{2(n\sigma_n)^2 t^2}\right\}$$
$$= \mathcal{O}\left(\frac{n(n\sigma_n)^2}{k^2}\exp\left\{-\frac{1}{2(1+\varepsilon)}\frac{(k/2)^3(1-\gamma\rho t^2)^2}{2(n\sigma_n)^2 t^2}\right\}\right).$$

The expression on the right-hand side is, when $n$ is large,

$$= \mathcal{O}\left(\frac{(n\sigma_n^{2/5})^{5/3}}{(\log(n\sigma_n^{2/5}))^{2/3}} \exp\left\{-\frac{(1+\varepsilon)^2 c_2(t)^3 (1-\gamma\rho t^2)^2 \log(n\sigma_n^{2/5})}{2^5 t^2}\right\}\right)$$

$$= \mathcal{O}\left(\frac{(n\sigma_n^{2/5})^{5/3}}{(\log(n\sigma_n^{2/5}))^{2/3}} \exp\left\{-\frac{5}{3}(1+\varepsilon)^2 \log(n\sigma_n^{2/5})\right\}\right),$$

which goes to 0 as $n \to \infty$. Hence,

$$\limsup_{n\to\infty} \frac{L_n(t)}{(n\sigma_n)^{2/3}(\log(n\sigma_n^{2/5}))^{1/3}} \leq (1+2\varepsilon)c_2(t) \qquad \text{in probability.}$$

This yields the desired upper bound in Theorem 3.5 for the lattice model.

The theorem is thus proved for the lattice model. It turns out that in the considered range of initial speeds, the fluctuations in initial positions have no significant influence upon the asymptotics of $L_n(t)$. Indeed, for arbitrary initial positions, (5.29) becomes

(5.33)
$$P_1 = \mathbb{P}\left(k^{-1/2}\left(S_j - \frac{j}{k}S_k\right) \geq \frac{k^{1/2}}{n\sigma_n t}\left(\frac{k}{2}(1-\gamma\rho t^2) + D_j\right)\frac{j}{k}\left(1 - \frac{j}{k}\right),\right.$$
$$\left.\varepsilon \leq \frac{j}{k} \leq 1 - \varepsilon\right),$$

where

$$D_j := \frac{1}{j}\sum_{i=1}^{j}(nX_i - i) - \frac{1}{k-j}\sum_{i=j+1}^{k}(nX_i - i).$$

Hence, for any $\varepsilon_1 > 0$, we have $P_1 \geq \widetilde{P}_1 - P_4$, where

$$\widetilde{P}_1 := \mathbb{P}\left(\min_{j \in J_0}(\bar{u}_{(0,j]} - \bar{u}_{(j,k]}) \geq \frac{(1+\varepsilon_1)k(1-\gamma\rho t^2)}{2n\sigma_n t}\right),$$

$$P_4 := \mathbb{P}\left(\max_{j \leq k}|D_j| \geq \frac{\varepsilon_1 k(1-\gamma\rho t^2)}{2}\right).$$

In the Poisson model, it follows from the large deviation principle that for some constant $c = c(\gamma, \rho, t, \varepsilon_1) \in (0, \infty)$,

(5.34) $$P_4 \leq \exp\{-ck\},$$

whereas from what we have proved for the lattice model, we know that $\widetilde{P}_1 \gg \exp\{-ck\}$. Therefore, the proof of the lower bound for the Poisson model goes through along the same lines as for the lattice model, with $\widetilde{P}_1$ in place of $P_1$.

The same happens with the upper bound, where $P_q$ should be replaced by

$$(5.35) \quad \widetilde{P}_q = \mathbb{P}\left(\frac{1}{\sqrt{2q}}\sum_{j=1}^{q}(u_{i+j} - u_{i+q+j}) \geq \frac{q^{1/2}}{\sqrt{2}n\sigma_n t}\left(\frac{q(1-\gamma\rho t^2)}{2} + D_q\right)\right).$$

Again, the exponential bound in (5.34) suffices to conclude the proof as for the lattice model.

The passage between the Poisson and the i.i.d. models via Fact 2.1 is straightforward.

5.4. *Proof of Proposition* 3.6 (*a sketch*). Few changes are needed with respect to the proof of Theorem 3.5, except that we have to provide alternative tools to those based on exponential moments. For the lower bound, we can replace the Komlós–Major–Tusnády estimate (5.31) by the Sakhanenko bound (see [13]) which states that under condition of the finiteness of the $p$th moment,

$$\mathbb{P}\left(\max_{1\leq j \leq k}|W_j - S_j| \geq r\right) \leq \frac{Ck}{r^p} \qquad \forall\, k \in \mathbb{N},\ \forall\, r > 0,$$

where $C > 0$ is an unimportant constant. For the upper bound, we can use, instead of Cramér's moderate deviation principle (Fact 5.4) in the estimate of $P_q$, the following result of Nagaev [11]: for $\alpha \in (0,1)$ and i.i.d. random variables $Y_1, \ldots, Y_q$ with mean zero, unit variance and finite $p$th moment,

$$\mathbb{P}\left(\sum_{j=1}^{q} Y_j \geq y\sqrt{q}\right) \leq C_{\alpha,p}[q^{1-p/2}\mathbb{E}|Y|^p y^{-p} + \exp\{-\alpha y^2/2\}]$$

$$\forall\, q \in \mathbb{N},\ \forall\, y > 0,$$

where $C_{\alpha,p} > 0$ is an unimportant constant. The rest of the proof is along the same lines.

5.5. *Proof of Theorem* 3.7 (*a sketch*). Under (3.10), the fluctuations of speeds and initial positions have comparable influence on the asymptotic behavior of $L_n(t)$, hence, we have to take both of them into account. Fortunately, the large deviation principle provides less precise (than in the preceding proofs), but still sufficient, estimates. We just outline how the estimation works for the upper bound for the Poisson model. The changes in the proof of the lower bound are similar.

Fix $M > 0$; let $q \geq M\log n$ and $B := \frac{1-\gamma\rho t^2}{4}$. We obtain for $P_q$ in (5.35) that for any $\varepsilon > 0$ and all large $n$,

$$P_q = \mathbb{P}\left(\sum_{j=1}^{q}(u_{i+j} - u_{i+q+j}) \geq \frac{q^2}{n\sigma_n t}\left(\frac{1-\gamma\rho t^2}{2} + \frac{D_q}{q}\right)\right)$$

$$\leq \mathbb{P}\left(\frac{|D_q|}{q} \geq B\right) + \mathbb{P}\left(\sum_{j=1}^{q}(u_{i+j} - u_{i+q+j}) \geq \frac{Mq(1-\varepsilon)}{ct}\frac{(1-\gamma\rho t^2)}{4}\right)$$

$$\leq e^{-(1+o(1))I_1(B)q} + e^{-(1+o(1))I_2(C)q},$$

where $C := \frac{M(1-\varepsilon)(1-\gamma\rho t^2)}{4ct}$, $I_1$ and $I_2$ are relevant large deviation functions. We take $M$ so large that $MI_1(B) > 1$ and $MI_2(C) > 1$. Then

$$\mathbb{P}(L_n(t) > 2M \log n) \leq n \sum_{q \geq M \log n} P_q \to 0,$$

and the upper bound follows.

**Acknowledgments.** Part of this work was done while the first named author was visiting the Université Paris VI in April 2003. We wish to thank the Laboratoire de Probabilités et Modèles Aléatoires (UMR 7599) for arranging this visit and for the hospitality. Our special thanks go to the referee and an Associate Editor for many valuable remarks.

FACULTY OF MATHEMATICS AND MECHANICS
ST. PETERSBURG STATE UNIVERSITY
198504 STARY PETERHOF
BIBLIOTECHNAYA PL. 2
RUSSIA
E-MAIL: lifts@mail.rcom.ru

LABORATOIRE DE PROBABILITÉS
ET MODÈLES ALÉATOIRES
UNIVERSITÉ PARIS VI
4 PLACE JUSSIEU
F-75252 PARIS CEDEX 05
FRANCE
E-MAIL: zhan@proba.jussieu.fr